\documentstyle[twocolumn,aps,prl,subfigure,epsfig]{revtex}
\def\meanchi{\overline{\chi^2}}
\def\brameanchi{\langle \meanchi \rangle}
\newcommand{\bi}[1]{\mbox{\boldmath $#1$}}
\begin{document}
\twocolumn[\hsize\textwidth\columnwidth\hsize\csname
@twocolumnfalse\endcsname

\title{Phase Mapping as a Powerful Diagnostic of Primordial Non-Gaussianity}
\author{Lung-Yih Chiang, Pavel Naselsky}
\address{Theoretical Astrophysics Center, Juliane Maries Vej 30,
DK-2100,  Copenhagen, Denmark}
\author{Peter Coles}
\address{School of Physics \& Astronomy, University of
Nottingham, University Park, Nottingham NG7 2RD, United Kingdom}

\date{\today}
\maketitle
\begin{abstract}
The identification and extraction of non-Gaussian signals is one
of the main cosmological challenges facing future experimental
measurements of the cosmic microwave background temperature
pattern. We present a generalized statistical measure based on a
novel technique representation of Fourier phases using the return
map. We show that this method is both robust and powerful in
comparison, for example, with morphological measures.
\end{abstract}
]

\section{Introduction}
One of the key challenges facing modern cosmology is the complete
statistical characterization of the primordial density
fluctuations believed to be the seeds of the large-scale structure
of the Universe we see today. According to the prevailing
orthodoxy, these initial perturbations were generated as quantum
phenomena during an inflationary epoch \cite{inflation}. If this
is the case they should display Gaussian statistics. Among other
potentially testable consequences of this is that the small
angular variations of the cosmic microwave background (CMB) sky
should have the same statistical form \cite{bbks,be}. Testing the
hypothesis of primordial Gaussianity using maps of the CMB thus
provides an opportunity to make experimental tests of ideas about
the physics of the early universe. It is also necessary to devise
robust statistical descriptors  for the identification of
foreground contaminations and systematic instrumental artifacts.

Searches for non-Gaussian (NG) signals within the COBE-DMR data
have yielded positive detections that are at least partly
explained by systematics \cite{nongaussian}. The need to devise
more powerful statistical probes has been recognized and acted
upon, with key ideas focusing on the bispectrum \cite{bispectrum}
and higher-order polyspectra \cite{trispectrum,second}. Some of
these methods have been developed as far as practical
implementation on real data sets \cite{bisp_measured}. These
techniques are sensitive only to particular types of NG signal
\cite{wc}, so in this paper we present a new method that is both
extremely general and extremely robust.

\section{Phase Mapping}
One of the most basic properties of a Gaussian random field (GRF)
\cite{bbks,be} is that the Fourier phases are random and
uniformly distributed between 0 and $2\pi$. The Central Limit
Theorem virtually guarantees that a superposition of a large
number of Fourier modes with random phases will be Gaussian, so
the random-phase hypothesis even serves as a definition of a form
of Gaussianity.

Testing the randomness of measured phases therefore provides a
direct diagnostic of the statistical form of a fluctuation field.
There are, however, some problems in using phases as a statistic.
The foremost is that they are circular variables, only defined
modulo $2\pi$. Traditional measures of association, such as
covariances of the form $\langle \phi_{\bi k} \phi_{\bi k'}
\rangle$, are based on the assumption that the related measure is
linear and are therefore inappropriate. Another problem is that
phases are direct indicators of the morphology and location of
specific features \cite{cc1}, so are themselves not
translational-invariant. As the first attempt to use phases as the
basis of a statistical test, ref. \cite{cc2} used a quantity
formed by taking the phase differences of neighboring Fourier
modes in $k$-space. This idea is the first step towards our new
``phase-mapping'' technique \cite{mapping}. We counter the problem
of the circular nature of the phases by applying phases on to a
``return map'', as follows.

For a set of phases $\phi_k$ from the Fourier transform of a
one-dimensional process, a possible return map is a plot of
$\phi_k$ against $\phi_{k+1}$ (neighboring phases) for all
available pairs $(\phi_k, \phi_{k+1})$ up to the Nyquist frequency
$k_{N/2}$ where $N$ is the length of the process \cite{cc1}. When
the phases are random this will be a scatter plot with points
distributed randomly within the bounded square of both axes
$[0,2\pi]$.  For a realization of two or more dimensions, we can
extend the return mapping between neighboring phases to any fixed
scale $\Delta {\bi k} \equiv (\Delta k_x, \Delta k_y) \equiv
(m,n)$ (so mapping of neighboring phases is simply the mapping for
$|\Delta {\bi k}|=1$). Therefore, this method forces to test
randomness between all pairs of $\phi_{\bi k}$ and $\phi_{{\bi
k}+\Delta{\bi k}}$, for all fixed $\Delta {\bi k}$ scales in the
Fourier domain. What results is a ``super-map'', each pixel of
which represents a return map with a fixed scale $\Delta {\bi k}$.

\section{Statistical Analysis of Phase Maps}
Having rendered the phase information in a pattern onto the phase
map, we have to consider the construction of a statistical test
using it. Our null hypothesis at each $\Delta{\bi k}$ case is a
random (Poisson) distribution on the mapping square, so the
randomness of phases is  tested through $2 \times m_{\max} \times
n_{\max}$ squares, where $m_{\max}$ and $n_{\max}$ are the maximal
scale in $k_x$ and $k_y$ direction, respectively.  For simplicity and
computational ease \cite{mapping}, we apply a test (similar to the
$\chi^2$ test) of the uniformity of the pixelized mapped phases for
different scales $\Delta {\bi k}$ as follows. A mean $\chi^2$
statistic is defined as
\begin{equation}
\meanchi=\frac{1}{M}\sum_{i,j}\frac{\big[ p(i,j)-\overline
p\big]^2}{\overline p} \equiv \sigma^2 \overline p,
\label{eq:meanchisquare}
\end{equation}
where $M$ is the number of pixels on the return map, $\overline p$
is the mean value for each pixel, and  $\sigma^2 = (1/M)
\sum_{i,j} \left[p(i,j)/\overline p- 1\right]^2$ is the variance
of the contrast of the return map.

The return mapping of random phases will result in a Poisson
distribution over all the map squares. For a Poisson distribution
of integer $p(i,j)$, its variance $\langle(\Delta p)^2
\rangle_{\rm Poisson}= \sum_{p=0} ^{\infty} (p-\overline p)^2
(\overline p^{p} e^{-\overline p} / p! ) =\overline p $. Hence in
this case $\meanchi_{\rm Poisson}=1$, and from
Eq.~(\ref{eq:meanchisquare}), $\sigma_{\rm Poisson}^2 =
(\overline p)^{-1}$. One could use this result to test the return
map for consistency with pure discreteness fluctuations. However,
such a test is not so useful as it does not probe the spatial
arrangement of the pattern. The individual return
maps tend to display particular patterns to which this would
not be sensitive. We could, for example, populate the pixels with
a Poisson distribution of counts but then rearrange them
arbitrarily over the grid without changing the statistical
properties mentioned above. One could sort the counts in ascending
order, for example, producing a clearly non-uniform distribution
that has the same statistics as the starting case.

In order to apply this method more usefully we have to take into
account the spatial correlations, which we can do using smoothing.
For a 2D square $p(i,j) \equiv p({\bi x})$, we use the form
%\begin{equation}
%p({\bf x},R)= (\sqrt{2\pi}R)^{-2} \int d^{2}{\bf x^{'}} \: p({\bf
% x^{'}}) \: \exp\left(-\frac{|{\bf x}-{\bf x^{'}}|^{2}}{2R^{2}}\right),
%\label{eq:window}
%\end{equation}
\begin{equation}
p({\bi x},R)=  \int \frac{d^{2}{\bi x^{'}}}{2 \pi R^{2}} \: p({\bi
 x^{'}}) \: \exp\left(-\frac{|{\bi x}-{\bi x^{'}}|^{2}}{2R^{2}}\right),
\label{eq:window}
\end{equation}
where $p({\bi x},R)$ is the smoothed square and $R$ is the smoothing
scale.

To obtain the expectation value of $\meanchi$ statistic of GRFs, we
calculate first the variance of the contrast of the smoothed
square, using a Gaussian window filter as in
Eq.~(\ref{eq:window}). For a 2D field with power spectrum
$P(k)=Ak^{n}$, the variance of the contrast of the smoothed square
\cite{bbks} is  $\sigma^2(R) = (A/4 \pi) \Gamma(n+2/2) R^{-(n+2)}$,
%\begin{eqnarray}
%\sigma_R^2 & = & \frac{1}{(2\pi)^2}\int^{\infty}_{0} Ak^n
%\exp(-k^2R^2) 2\pi k \, dk \\
%& = & \frac{A}{4 \pi} \Gamma\left(\frac{n+2}{2}\right) R^{-(n+2)},
%\end{eqnarray}
%\begin{equation}
%\sigma_R^2 = \frac{A}{4 \pi} \Gamma\left(\frac{n+2}{2}\right) R^{-(n+2)},
%\end{equation}
where $\Gamma(n)$ is the Euler Gamma function and $R$ is the
smoothing scale in Eq.~(\ref{eq:window}).  A Poisson
distribution has a power spectrum which is independent of wave number,
i.e., its spectral index $n=0$. So originally the variance of the contrast
$\sigma_{\rm Poisson}^2 = A = (\overline p)^{-1}$, after smoothing,
becomes $\sigma^2_{\rm Poisson}(R) = (4 \pi R^2 \overline
p)^{-1}$. According to Eq.~(\ref{eq:meanchisquare}), we can obtain the
$\brameanchi$ from the variance of the smoothed return map and the
$\overline p$, so for an ensemble of Poisson distributions,
the expectation mean $\chi^2$
\begin{equation}
\brameanchi = \frac{1}{4 \pi R^2}. \label{eq:theoreticalvalue}
\end{equation}

Thus, any smoothed return map of phase pairs at a  fixed $\Delta
{\bi k}$ that generates a $\meanchi$ value much higher than $(4
\pi R^2)^{-1}$ should be interpreted as a diagnostic of phase
coupling at that particular scale in phase space. To apply phase
mapping technique as a NG test, the null hypothesis requires that
phase mapping for any fixed scale $\Delta {\bi k}\equiv(m,n)$
should result in Poisson distribution in a return map. Therefore,
for a Gaussian realization, we have an ensemble of $2 \times
m_{\max} \times n_{\max}$ smoothed return squares which are
subject to statistical fluctuations around $\brameanchi$. Although
phase mapping extracts all the information between phase pairs, we
have to know whether the value obtained from $\meanchi$ statistic
is due to non-randomness or statistical fluctuation. As the
$\meanchi$ statistic is related to the `variance' of the smoothed
Poisson-distributed return map, the problem now is equivalent to
finding the distribution of sampling variance from normal samples.
The statistical significance of the departure of a measured value
of $\meanchi$ can be assessed using standard techniques for the
sampling distribution of the variance in normal samples. When the
sampling number is large enough, it behaves as Gaussian. We have
$M$ independent variables $p_i-\overline p/ \sqrt{\overline p}$,
each of which is distributed normally with zero mean and  variance
$\brameanchi$. We are now looking for the distribution of sampling
variance $w \equiv \meanchi = (1/M)\sum_i (p_i-\overline
p)^2/\overline p$, which can be obtained as \cite{kendall},
\begin{equation}
dF=\frac{(M/2\brameanchi)^{(M-1)/2}}{\Gamma[(M-1)/2]}
\exp \left(-\frac{M w}{2\brameanchi }\right) w^{(M-3)/2} dw.
\end{equation}
Therefore, the variance of the sampling $\meanchi$ measurements
($M \gg 1$) is $\Sigma^2 =\int_0^{\infty}(w- \brameanchi)^2 dF =
\brameanchi^2/(M/2)$. The dependence of $\Sigma^2$ on $R$ can be
obtained through Eq.~(\ref{eq:theoreticalvalue}), and the change of
the {\it effective} pixel number $M$ to $(\pi/2)^2 M/ (4 \pi R^2)$ due
to smoothing. Therefore,
\begin{equation}
\Sigma^2=\frac{1}{\pi^3 R^2 (M/2)}.
\end{equation}

We can examine NG signals in two ways. A pixel in the
``super-map'' which corresponds to $\nu\Sigma$ with $\nu \simeq
5$, say, would be extremely unlikely if the null hypothesis held.
As well as easily identifying such hotter pixels, we can also
retrieve the corresponding $\Delta {\bi k}$ scale of the signal
from the map. One could look at modes flagged in this way more
carefully using other statistics if necessary. On the other hand,
we can also inspect the global property of the distribution curve
with $\brameanchi$ and $\Sigma$ when the number of $2 \times
m_{\max} \times n_{\max}$ is large. Even when all signals are
below $5 \Sigma$, the underlying obscure NG part could cause the
distribution curve to skew, albeit no signal is above $5\Sigma$
due to possible 'Gaussianization' by the noise.

\section{Application to simulated CMB maps}
We apply this method to a simulated NG CMB map (Fig.1(a)), which is a
realization of $12.8 \deg^2$ CMB anisotropies due to the
Kaiser-Stebbins effect from cosmic strings \cite{bouchet}. We will
test the non-Gaussianity by applying our method not only directly to
the original manifestly NG map (S), but also to the combined map of
itself plus Gaussian white noise (N) with 5 different fluctuation
levels, as NG part may be embedded in white noise. The Gaussian white
noise level are chosen with {\it rms} ratio ${\rm S/N}=8$, 4, 2, 1 and
1/2.

\begin{figure}[t]
\centering 
\epsfxsize 5 cm
\subfigure[]{\epsfbox{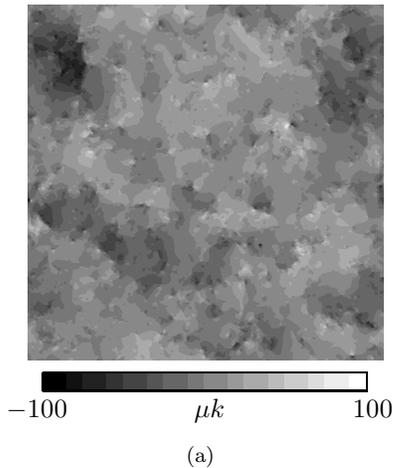}}
\put(-15,-5){$100$} \put(-146,-5){$-100$} \put(-75,-5){$\mu k$}
\put(-132,5){\framebox(122,6.5)[t]} \\ 
\epsfxsize 9cm
\subfigure[]{\epsfbox{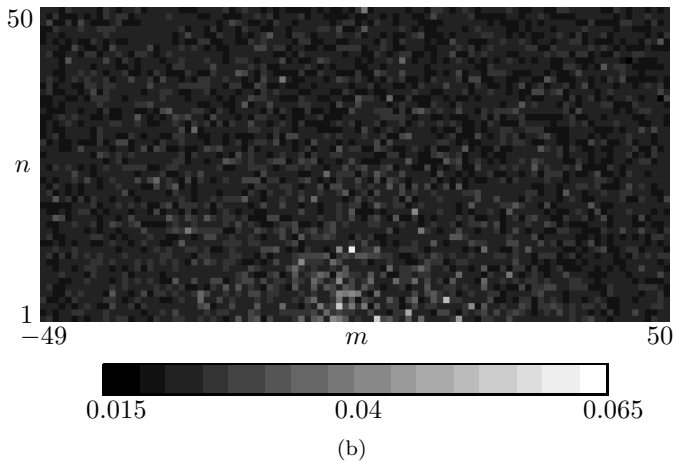}} \put(-18.,20.){$50$}
\put(-255.,28){$1$} \put(-132.,20.){$m$} \put(-255.,20.){$-49$}
\put(-42.,-8.){$0.065$} \put(-136.,-8.){$0.04$}
\put(-230.,-8.){$0.015$} \put(-260,140){$50$} \put(-257,85){$n$}
\put(-138.,154){} \put(-223,1){\framebox(190,11)[t]} 
\caption{The
temperature map ($\mu k$) and the ``super-map'', in which each point
represents the $\meanchi$ value for that $(m,n)$ scale. The histogram
of the latter is shown in Fig.~\ref{histo} (a).}
\end{figure}

From the Fourier transform of a $N^2$-mesh realization, we have
$N^2/2$ valid modes: $k_x \in [-N/2+1, N/2]$ and
$k_y \in [1, N/2]$. In general, the phase basis used to extract
NG signals is flexible with the constraint $|{\bi
k}|_{\max}+|\Delta {\bi k}|_{\max} \le k_{\rm Nyquist}$. We have
taken phases from the inner quarter of valid modes as a basis,
i.e., $k_x \in [-N/4+1, N/4]$, and $k_y \in [1, N/4]$.
As the Gaussian noise normally dominates the power spectrum at
large $k$'s, we can opt to avoid that part of `Gaussianized'
phases by shrinking the upper limit $|{\bi k}|_{\max}$ of the
phase base (not the upper limit of $(m,n)$).
%This shrinking, however, does not extract effectively more than
%otherwise in the string map, as in this case the phase coupling
%occurs at large $k$.

\begin{figure}
\centering
\epsfig{file=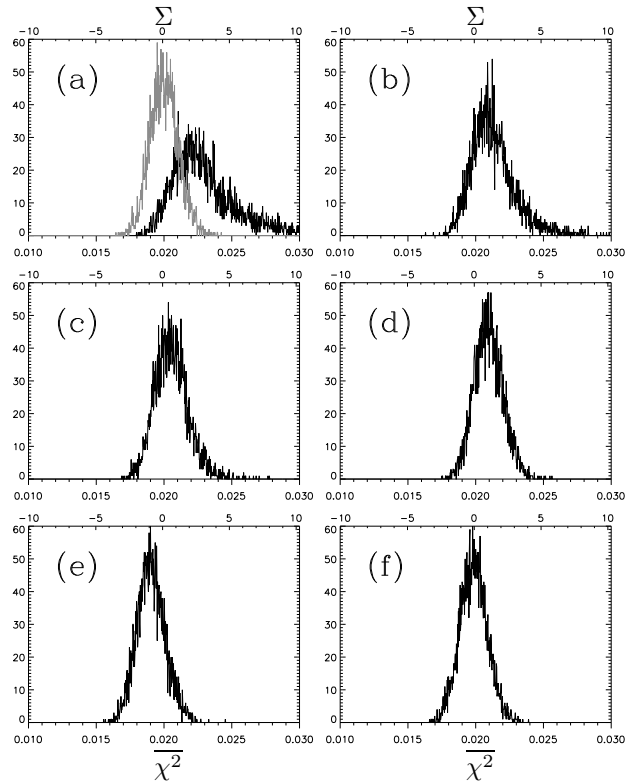,width=9cm}
\put(-70,13){$\meanchi$}
\put(-188,13){$\meanchi$}
\put(-70,297){$\Sigma$}
\put(-188,297){$\Sigma$}
\caption{The histogram of $\meanchi$ statistics. Panel (a) is from
the pure temperature map caused by cosmic strings of
Kaiser-Stebbins effect. For comparison, the gray curve is from
a realization of (random-phase) GRF. Panel (b), (c), (d), (e) and (f)
are the histograms for the combined map of CMB plus
Gaussian white noise with {\it rms} ratio ${\rm S/N}=8$, 4, 2, 1 and
$1/2$, respectively. The upper horizontal axis is annotated in terms
of the theoretical value $\Sigma$ of GRFs with origin set at
$\brameanchi =(4 \pi R^2)^{-1}$. The smoothing scale on the $M=128^2$
pixelized return map is $R=2$. Therefore we regard those above $5
\Sigma$ as against the null hypothesis, thus manifestation of phase
coupling. We can see that phase mapping is able to detect considerable
signals above $5 \Sigma$ for the added Gaussian noise ${\rm S/N}=4$
and some for ${\rm S/N}=2$. Note that albeit no signals above $5
\Sigma$ in (e), the median is shifted away from the $\brameanchi$
value, which is also NG manifestation.}
\label{histo}
\end{figure}

In Fig.~\ref{histo} we show the histogram of the measured
$\meanchi$ statistics. Those above $5\Sigma$ are caused by phase
coupling of some $\Delta{\bi k}$, which can be viewed in the
$(m,n)$-map as in Fig.1 (b). As a NG test, we are interested only in
the amplitude of phase coupling in terms of $\meanchi$, whatever the
$\Delta{\bi k}$.

\section{Comparison to the Analysis of Minkowski Functionals}
\begin{figure}
\centering
\epsfig{file=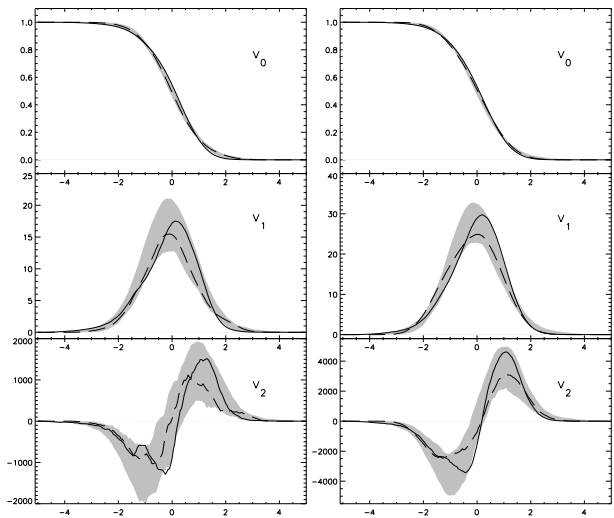,width=9cm} \caption{Minkowski
functionals for the combined temperature map with ${\rm S/N}=4$
(left column) and ${\rm S/N}=2$ (right) in solid curves
against threshold (in terms of {\it rms}). From top downwards are
the area, circumference and the Euler characteristic. The shaded
areas are the possible ranges for each functional from $10^3$
realizations of the same power spectrum (as the noise--added CMB map)
but random phases and the dash curves are from one of the $10^3$ Gaussian
realizations.} \label{mf}
\end{figure}

We now compare this method with a more standard morphological
analysis of temperature contours, based on Minkowski functionals
(MFs). The 3 MFs in 2D are the area ($V_0$), circumference
($V_1$), and  the Euler characteristic ($V_2$) of excursion
regions \cite{mf}. The excursion region is the region on the sky
map above a certain threshold amplitude \cite{coles}. In
Fig.~\ref{mf} we show the 3 MFs for the noise-contaminated cases:
CMB plus noise with ${\rm S/N}=4$ and 2 shown by solid curves. The
shaded areas are from $10^3$ realizations of the same power
spectrum of the corresponding noise--contaminated CMB map but with
random phases, which thus represents the statistical fluctuations
of GRFs. We avoid quoting $1\sigma$ around the mean MFs as the
distribution function for each threshold is not normal (e.g. for
high thresholds $\nu$, $V_2 \propto \nu \;e^{-\nu^2/2}$
\cite{be}). The case of ${\rm S/N=4}$ is out of the shaded area, a
definite NG manifestation, but ${\rm S/N=2}$ is right on the edge.
The dashed curves are taken from one Gaussian realization in order
to show the variation of the MFs and the ambiguity that would
result.

Other types of NG signals {\it localized} in harmonic space,
e.g., artifacts from systematics such as slowly--rotating elliptical
beam of the {\sc planck} mission will produce NG signals around
$\ell_{\rm pix}$, which is more easily detected with harmonic analyses than
real-space ones. By ``localized in harmonic space'' we mean the
phases correlate mainly from large ${\bi k}$. The contrary example is
the simulations of spectral index $n=2$ \cite{mapping}, where phases
correlate first at large $\Delta{\bi k}$ from all ${\bi k}$.

These results show that the phase-mapping technique is able to
detect certain forms of non-Gaussianity even in the presence of
noise, more effectively than the MF approach. The theoretical ground
for phase mapping we have developed as a powerful NG diagnostic can be
directly applied not only on spherical analysis, but also on
time-ordered scans of the upcoming {\sc planck} mission.

This paper was supported in part by Danmarks Grundforskningsfond
through its support for the establishment of TAC by grants RFBR
17625. PC acknowledges support from PPARC. We are grateful to
A. A. Starobinsky for useful discussions.

\def\jnl#1#2#3#4#5{\hang{#1, #3, {\bf #4}, #5 (#2).}}
\def\jnlibid#1#2#3#4#5{\hang{#1, {\it ibid.} {\bf #4}, #5 (#2)}}
\def\prep#1#2#3#4{\hang{#1, #4.}}
\def\proc#1#2#3#4#5#6{{#1 [#2], in {\it #4\/}, #5, eds.\ (#6).}}
\def\book#1#2#3#4{\hang{#1, {\it #3\/} (#4, #2).}}

\def\prl{Phys.\ Rev.\ Lett.}
\def\pl{Phys.\ Lett. }
\def\np{Nucl.\ Phys. }
\def\rmp{Rev.\ Mod.\ Phys.}
\def\cmp{Comm.\ Math.\ Phys.}
\def\mpl{Mod.\ Phys.\ Lett.}
\def\apj{Astrophys.\ J.}
\def\apjl{Astrophys.\ J.\ Lett.}
\def\aa{Astron.\ Astrophys.}
\def\cqg{Class.\ Quant.\ Grav.}
\def\grg{Gen.\ Rel.\ Grav.}
\def\mn{Mon.\ Not.\ R.\ Astron.\ Soc.}
\def\nature{Nature (London)}
\def\cupress{Cambridge University Press}
\def\pup{Princeton University Press}
\def\wss{World Scientific, Singapore}
\def\oup{Oxford University Press}

\end{document}